\documentclass[%
superscriptaddress,
 amsmath,amssymb,
 aps,
]{revtex4-2}
\usepackage{geometry}
\usepackage{subcaption}
\usepackage{graphicx}
\usepackage{dcolumn}
\usepackage{color}
\usepackage{comment}

 \global\long\def\halfsize{0.45\columnwidth}
\usepackage{bm}
\usepackage{comment} 
\usepackage{hyperref}
\hypersetup{colorlinks=true, citecolor=blue, urlcolor=blue, linkcolor=blue}
\begin{document}
\title{Initial conditions for Starobinsky Inflation with a positive spatial curvature}
\author{Daniel M\"uller} 
\email{dmuller@unb.br}
\affiliation{Instituto de F\'{i}sica, Universidade de Bras\'{i}lia, Caixa Postal
04455, 70919-970 Bras\'{i}lia, Brazil}
\author{Alexey Toporensky} 
\email{atopor@rambler.ru}
\affiliation{Kazan Federal University, Kazan 420008, Republic of Tatarstan, Russian Federation}
\affiliation{Sternberg Astronomical Institute, Moscow University, Moscow 119991, Russian Federation}
\date{\today}

\begin{abstract}
    We have found numerically initial conditions in the $(R, H)$ plane leading to a successful Starobinsky inflation in $R+R^2$ gravity for a isotropic metrics with positive spatial curvature. 
    Trajectories can reach inflation regime either directly or going through a bounce, and even recollapse  followed by a bounce. Our numerical plots indicate that ``good" initial conditions exist even
    for big initial spatial curvature, however, we argue that such a trajectory must cross a region of rather big $R$ or $H$. This means that the range of viability of $R+R^2$ theory in the $(R,H)$ plane
    directly affect the question of viability of Starobinsky inflation for a positive spatial curvature isotropic Universe.
\end{abstract}
\maketitle
\section{Introduction}
During the inflationary scenario, the gravitational field is sufficiently strong that quadratic gravity could be the correct gravitational theory, in an perturbative effective sense. Quadratic gravity is not new and the first to investigate it was H. Weyl in 1918 \cite{weyl1918gravitation}. The subject emerged  again in the 60s 70s \cite{buchdahl1962gravitational,tomita1978anisotropic,Buchdahl:1983zz,Gurovich:1979xg} and more recently \cite{Berkin:1991nb,Cotsakis:1997ck,Cotsakis:2007un,Miritzis:2003eu,Miritzis:2007yn,Barrow:2006xb,Barrow:2009gx,Barrow:2005qv,Carloni:2007br,Toporensky:2016kss,vitenti2006numerical,Salvio:2015kka,Castardelli_dos_Reis_2019}, for a review specific to quadratic gravity see \cite{schmidt2007fourth}. Quadratic gravity gives rise to a number of solutions impossible in General relativity, such as stable static solution \cite{PhysRevD.76.084005,PhysRevD.78.044011,Goheer_2009,Muller:2021qzz} or ``dark radiation" vacuum past attractor \cite{Muller:2017nxg}. One of such regimes is a vacuum inflation which does not need any additional matter
to realize found many years ago by Starobinsky \cite{starobinsky1980new}. 

Indeed, Planck's CMB observations \cite{Ade:2015lrj} indicate that Starobinsky inflation \cite{starobinsky1980new} which is a particular case of quadratic theory, is currently the best inflation model, which excellently  fits the data. That is why a detailed investigation of initial conditions leading to this regime is important.
Previous works have investigated  this problem for a flat Universe, \cite{Mishra:2018dtg} and \cite{Mishra:2019ymr}. For a review of inflation in gravitational theories see \cite{Nojiri_2017}.


Positive curvature effects however may be important as the known qualitative argument states. Theory of Starobinsky inflation is equivalent to a GR inflation theory with a scalar field (so called Einstein frame) having a special type of
potential tending to a constant value while field $\phi \to \infty$. This means that if a classical evolution starts from a big (for example, Planck) energy, the most part of it is in the form of kinetic energy of the scalar field 
which is known to decay more rapidly than the curvature term. This could cause the recollapse of the Universe instead of inflation \cite{ijjas2013inflationary} (see, however, the discussion in\cite{Gorbunov:2014ewa}) . Though this reasoning can not be directly applied to Jordan frame picture because
it is hard to define unambiguous kinetic term, and the effective Planck energy is no longer a constant, the role of positive spatial curvature requires careful investigation.

This work specializes to Jordan frame which is considered the physical one, where intervals are determined directly with the definition of metric with the absence of conformal factors, the equivalence principle is directly satisfied. Also, it is in the Jordan frame that the Ruzmaikina-Ruzmaikin solution arises as the asymptotic solution connected to Starobinsky's inflation \cite{ruzmaikina1970quadratic}.

We consider throughout ``good"  inflation as $~60$ e-folds and investigate initial conditions numerically for inflation.
Integration is not halted at the point of recollapse (if it exists for a particular trajectory) since, as it is presented below, good inflation can follow after a recollapse followed by a bounce.


\section{Equations of motion} 
We consider $R+R^2$ vacuum gravity in $4$ dimensions
\begin{align}
L_g=\frac{1}{16G\pi}\left\{\beta R^2 +R\right\}.
\label{acao}
\end{align}
Metric variations result in the field equations
\begin{equation}
G_{ab}+\beta H_{\: ab}^{(1)}=0,\label{eq.campo}
\end{equation}
where 
\begin{eqnarray*}
 &  & G_{ab}=R_{ab}-\frac{1}{2}g_{ab}R,\\
 &  & H_{ab}^{(1)}=-\frac{1}{2}g_{ab}R^{2}+2RR_{ab}+2g_{ab}\square R-2\nabla_a\nabla_bR.
\end{eqnarray*}

In the following the appropriate line element for spatial $S^3$ is considered 
\begin{align}
ds^2=-dt^2+e^{2a}(d\psi^2+\sin(\psi)^2(d\theta^2+\sin(\theta)^2d\phi^2),
\end{align}
with scale factor $\mathcal{R}=e^a$. Of course in this setting, the Hubble parameter $H$ is given by
\[
H=\frac{\dot{\mathcal{R}}}{\mathcal{R}}=\dot{a},
\]
while Ricci scalar reads 
\[
R=6\left(\ddot{a}+2\dot{a}^2+\frac{1}{e^{2a}}\right).
\]
The $00$ equation from \eqref{eq.campo} 
\begin{align}
2\beta\dddot{a}\dot{a}+6\beta\ddot{a}\dot{a}^{2}-\beta\ddot{a}^{2}+\frac{\beta}{e^{4a}}+\frac{1}{6}\dot{a}^{2}-2\beta\frac{\dot{a}^{2}}{e^{2a}}+\frac{1}{6e^{2a}}=0
\label{eq-00}
\end{align}
is a constraint which is preserved on time evolution, while the $11$ equation from \eqref{eq.campo} contains the dynamics
\begin{align}
  -12\beta\dddot{a}\dot{a}-18\beta\ddot{a}\dot{a}^{2}-2\beta\ddddot{a}-9\beta\ddot{a}^{2}+\frac{\beta }{e^{4a}}
  -\frac{1}{3}\ddot{a}-\frac{1}{2}\dot{a}^{2}+\frac{4\beta\ddot{a}}{e^{2a}}+\frac{2\beta\dot{a}^{2}}{e^{2a}}-\frac{1}{6e^{2a}}=0.
 \label{eq-11}
\end{align}

As it is  well known, the Ruzmaikina-Ruzmaikin solution for Starobinsky inflation  is an asymptotic and obtained directly from \eqref{eq-00} when both $a\rightarrow \infty$ and $\dot{a}>>\ddot{a}>>\dddot{a}$
\begin{align}
    & \dot{a}^2\left(6\beta \ddot{a}+\frac{1}{6}\right)\sim 0 \nonumber\\
    &\dot{a}=H_0-\frac{t}{36\beta}.
    \label{RRS}
\end{align}
For more detailed treatment of this inflationary solution from the viewpoint of dynamical system theory see \cite{Alho:2016gzi}. As for the full dynamical equation (\ref{eq-11}) in the case of positive spatial 
curvature, it admits an exact solution in the form of power series \cite{Paliathanasis:2017apr}. The existence of such solution would indicate that the dynamics is regular. Indeed, in our numerical work (see below) we have
found that all boundaries of basins of attraction are regular and no chaotic behavior have been detected.

Besides this, as was also discussed in \cite{Mishra:2019ymr} there is a lower critical value for the Riemann scalar 
\begin{align}
    & R>-\frac{1}{2\beta}\simeq -3.381\times 10^{-10}, \label{critic_R}
\end{align}
for realistic value of $\beta=1.305\times 10^9$ in Planck mass units \cite{Ade:2015lrj}. For smaller Ricci scalars, there is no Einstein frame defined for the theory. 
\begin{figure}[h]
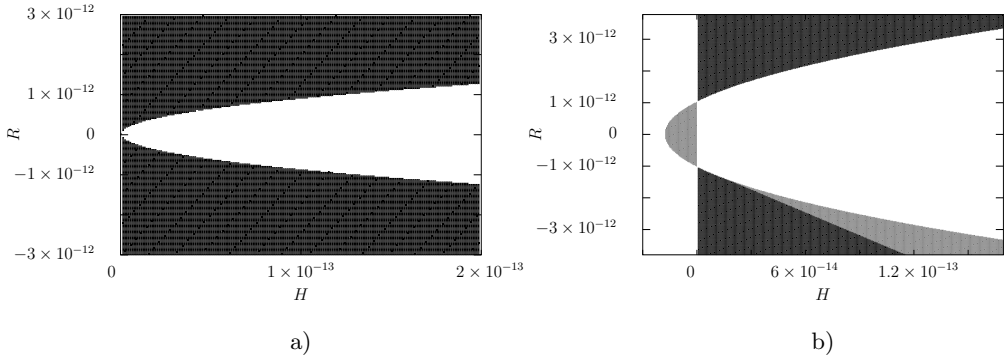

\centering
\begin{tabular}{c c} 
       \resizebox{\halfsize}{!}{\input{fig1}} 
       & \resizebox{\halfsize}{!}{\input{fig2}}\\
      a) & b) 
    \end{tabular}
\caption{a) Basin for very small spatial curvature with initial scale factor $\mathcal{R}=e^{24}$. This picture looks almost the same as for exactly zero spatial curvature presented in Fig.6a) of \cite{Mishra:2019ymr}.  b) The same for  initially bigger spatial curvature for scale factor $\mathcal{R}=e^{18}$. Black points mark initial conditions with sufficient inflation and no sign change in $H$, the Hubble parameter. Grey points are initial conditions also with sufficient inflation, that present at least one sign change in $H$.}	
\label{small_curv}
\end{figure}
\begin{figure}[h]
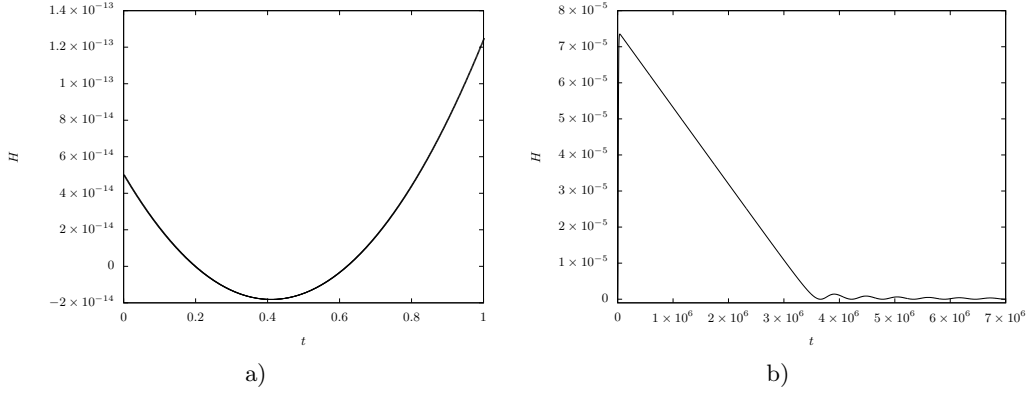

\centering
\begin{tabular}{c c} 
       \resizebox{\halfsize}{!}{\input{fig3}} 
       & \resizebox{\halfsize}{!}{\input{fig4}}\\
      a) & b) 
    \end{tabular}
\caption{a) It is plotted a single orbit showing both recollapse and bounce. The initial condition $H=5\times 10^{-14}$ and Riemann scalar $R=-2\times 10^{-12}$ with initial scale factor $\mathcal{R}=e^{18}$ is in the grey basin region $R<0$ and $H>0$ of FIG. \ref{small_curv}b). b) Continuation of the orbit showing the asymptotically Ruzmaikina-Ruzmaikin behavior.}	
\label{LABEL1}
\end{figure}
\begin{figure}[h]
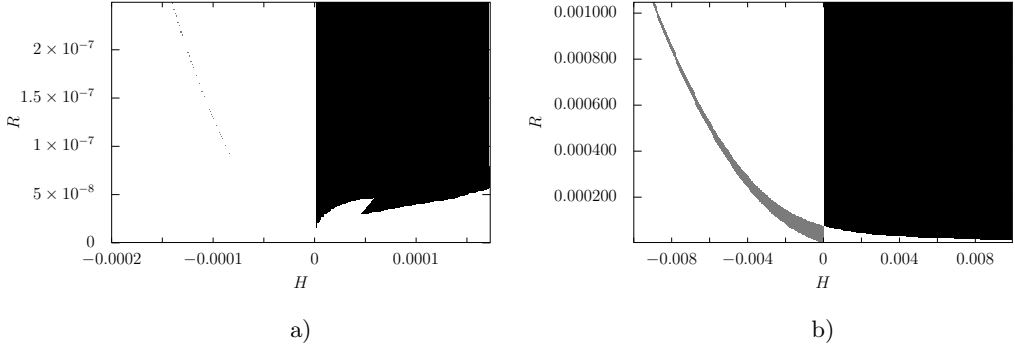

\centering
 \begin{center}
\begin{tabular}{c c} 
       \resizebox{\halfsize}{!}{\input{fig5.tex}} 
       & \resizebox{\halfsize}{!}{\input{fig6.tex}}\\
      a) & b) 
    \end{tabular}
    \end{center}
\caption{a) This basin is for initial spatial curvature specified by scale factor $\mathcal{R}=e^{10.43}$. b) Initial scale factor $\mathcal{R}=e^6$.}	
\label{fig3}
\end{figure}
\begin{figure}[h]
\centering
 \begin{center}
\begin{tabular}{c c} 
       \resizebox{\halfsize}{!}{\input{fig7.tex}} 
    &\resizebox{\halfsize}{!}{\includegraphics{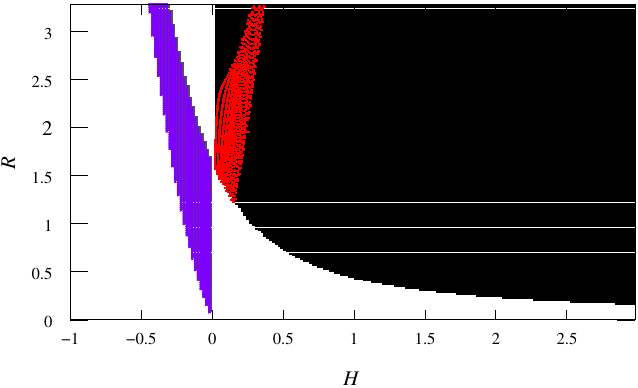}}\\
       a) & b)
    \end{tabular}
    \end{center}
\caption{a) This basin is for initial spatial curvature specified by scale factor $\mathcal{R}=e^1$. b) It is shown the return map for the $H<0$ region in a). All the points in violet are initial conditions in the surface $\mathcal{R}=e^1$ which return to this same surface $\mathcal{R}=e^1$, the return points are plotted in red. \label{fig_r}}	
\end{figure}
\begin{figure}[h]
\centering
 \begin{center}
\begin{tabular}{c c} 
       \resizebox{\halfsize}{!}{\includegraphics{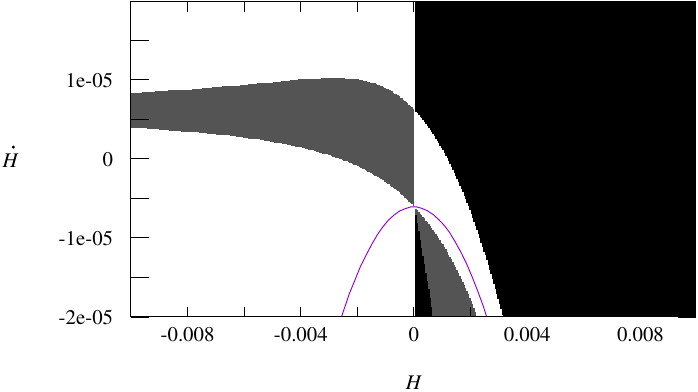}} 
    &\resizebox{\halfsize}{!}{\includegraphics{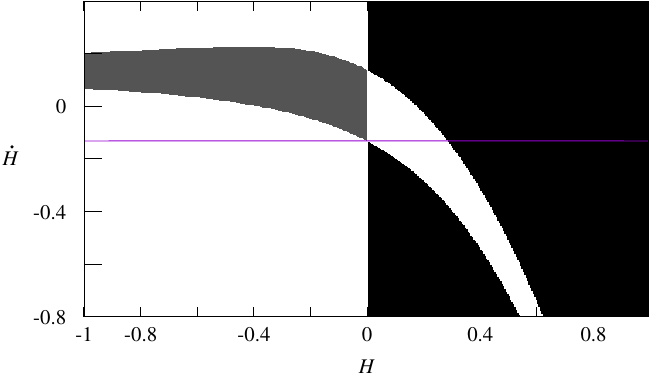}}\\
       a) & b)
    \end{tabular}
    \end{center}
\caption{a) The basin plot for $\mathcal{R}=e^6$ (the same as in  FIG. \ref{fig3} b)) shown in $(H,\dot{H})$ coordinates. The condition $R>-1/(2\beta)$ in \eqref{critic_R} is satisfied only above the violet line. b) 
The basin plot for $\mathcal{R}=e^1$ (the same as in  as in FIG. \ref{fig_r})  in coordinates $(H,\dot{H})$ and again the critical line \eqref{critic_R} is plotted in violet.\label{fig_H_dH}}	
\end{figure}

\section{Numerical results}

Our numerical results are presented in the form of a set of plots in which we mark the initial conditions leading to successful inflation by black, if a trajectory does not experience any bounce,
and by gray if a trajectory goes through a bounce on its way to inflation. Each plot is made
for a particular initial value of the scale factor. For comparison, almost zero curvature plot is reproduced in the FIG. \ref{small_curv} a). Small initial curvature (i.e. big initial scale factors) leads to a picture of the type of shown
in FIG. \ref{small_curv} b). An important difference with the FIG. \ref{small_curv} a) is that now inflation is possible even if start from an appropriate initial conditions with {\it negative} $H$.

Evidently, such initial conditions result in trajectories experiencing a bounce before reaching the inflationary regime. 
What is also remarkable is that most trajectories form the $R<0$ ``good" initial 
conditions zone first experience recollapse, then bounce  and finally reach inflation, as shown in FIG. \ref{LABEL1}. So that, recollapse
before inflation does not
necessary means that the trajectory is non-inflationary -- it can bounce back and restore the expansion.

As the initial curvature increases, the $R<0$ zone shifts to bigger absolute values of $R$ and already for $a \sim 10$ this zone goes down below the line of attractive gravity $R=-1/(2\beta)$, so that
starting from a negative $R$ becomes impossible for physically acceptable inflation trajectories. On the other side, FIG. \ref{fig3} a) shows that another zone of inflation trajectories appears, and this zone
is located in the $H<0$ half-plane. Possibly, it is present for even smaller initial curvatures, but since it is very narrow it is indistinguishable in our numerical plots. It becomes bigger with bigger
initial curvature (see FIG. \ref{fig3} b)). Trajectories from this gray zone have one bounce, after it they reach inflationary regime. 

The same value of initial scale factor $a \sim 10$ leads to important changes for the $H>0$ black zone. Zero curvature allows trajectories starting from arbitrary small $H$ and $R$ - the black zone touches the point $(0,0)$. Nonzero but small spatial curvature shifts the black zone to bigger values of $R$, 
but still bigger initial $H$ requires   bigger initial $R$ (see FIG. \ref{small_curv} a) and b)). For a large 
 spatial curvature this dependence is the opposite one (see FIG. \ref{fig3} b)). If the spatial curvature is taken from 
some transitional interval, like in FIG. \ref{fig3} a),  the lower boundary of the black zone shows some kind of an oscillatory behavior.

The described
picture appears to be qualitatively the same even for bigger curvature as it is shown in FIG. \ref{fig_r} a). We can see that the boundary of ``good" zone in general drifts up with increasing spatial curvature, so bigger values of $R$ are required for a fixed $H$ in order to get a successful inflation when starting from a big positive curvature. This can lead to problems of inflationary scenario since large initial $R$ or $H$ may be located beyond
the domain of validity of the described theory. Though there are no strict boundary (like a planckean density, since the effective Planck mass is not a constant any more), we can expect that  large initial values of Ricci curvature or the Hubble parameter may be inappropriate. For example, if we want to consider only trajectories with both $R$ and $H$ being smaller than $1$ in the units of the bare Planck mass, then appropriate initial
conditions still exist in the upper right corner of the square in question for $a=1$ presented in FIG. \ref{fig_r} a). It can be shown numerically that for $a=0.7$ all black points are located outside the square $0<R<1$, $0<H<1$. This means that an appropriate inflation can not be reached from such a large initial spatial curvature.
This situation can not be cured using $H<0$ ``good" zone. Though this zone touches the $(0,0)$ point, in is evident that a trajectory, leading to inflation,
while crossing the same spatial curvature {\it during expansion} should pass through $H>0$ ``good" zone. Fig.5 shows what part of the $H>0$ zone actually belong to bouncing trajectories. We can see that, in principle,
such trajectories could go through lower part of the corresponding zone, but as this whole zone drifts for bigger $R$ with smaller 
initial $a$,
 a bouncing trajectory starting near $(0,0)$ point at contraction will ultimately
go through rather high values of the Ricci curvature or Hubble parameter in order to reach the inflation if started from a large positive spatial curvature initial conditions.

As for the good initial conditions in the $H<0$ half-plane forming a prolonged ``tail", they correspond to small initial $\dot H$, as it can be seen from the FIG. \ref{fig_H_dH} representing the described zones
in the coordinates $(\dot H, H)$.

\section{Conclusions}

We have considered numerically the question of initial conditions leading to a successful inflation
in $R^2$ gravity. In this work we found that, in contrast to the zero spatial curvature case, the positive spatial curvature can cause both recollapse and bounce. On the other hand, since in this inflationary scenario the matter source is set to zero, a stable oscillatory state for the scale factor is not possible. In our plots we mark initial conditions leading to succesfull inflation as black points (or gray if a trajectory goes through a bounce), we do not discriminate between other outcomes marking all initial conditions which do not lead to a good inflation as white. 

Corresponding initial conditions have been presented as some zones in the $(R,H)$
plane. We have found that the configuration of these zones are of two general patterns corresponding to
low ($a>a_{cr}$) or high ($a<a_{cr}$) initial spatial curvature, where $a_{cr}$ is found to be of the 
order of $10$.

For low initial curvature the configuration of ``good" initial condition resembles 
those for zero curvature with two important differences. First, $H>0$, $R>0$ part of the zone becomes
separated from $(0,0)$ point by some finite interval. Second, a new compact zone appears in the $H<0$ part
of the plot. Trajectories from this new zone reach inflation after a bounce.

For large initial curvature instead of a compact zone in the range of negative $H$ we can see a long
(possibly, infinite) narrow tail. Initial conditions from this tail have small $\dot H$  as it can be seen from the plot in
the coordinates $(\dot H, H)$. As for zone in the $H>0$, $R>0$ part of the plot, it continues to shift
to bigger $R$, and its boundary being concave for low curvature becomes convex. Zone in $H>0$, $R<0$ part
disappears for physically admissible values of $R$.

What these plots tell us about? We can see that if there are no restrictions of the initial values of
$R$ and $H$, initial conditions leading to a successful inflation exists even for high initial 
spatial curvature and moreover no fine-tuning needed. On the other hand, very high values may be
inappropriate from the physical point of view since we can not be confident if the theory itself remains valid
for high 4-curvature regime. So that, any compact boundary of physically admissible initial conditions
in the $(R,H)$ plane would lead to disappearance of initial conditions, good for inflation, if initial
spatial curvature exceeds some value. This happens when black zone in $H>0$ half-plane shifts beyond
the physical boundary considered. For example, the condition $0<R<1$, $0<H<1$ in bare Plank units lead to disappearance
of a sufficient inflation for initial $a<0.7$.
Note that the existence of good initial conditions in the $H<0$
half-plane does not change the result, despite  the fact that such zone of initial conditions
 touches the $(0,0)$ point. The reason is that every point in the $H<0$ part of good initial conditions
zone  corresponds to a point in the $H>0$ part of good initial conditions zone - the point 
which  the trajectory in question goes through when crossing the same scale factor 
 at expansion after bounce.

\section*{ Acknowledgments} The work of  AT is  supported by RSF grant 21-12-00130. AT also thanks the Russian Government Program of Competitive Growth of Kazan Federal University.
 DM thanks FAPDF grant {\it visita t\'ecnica} no. $00193-00001537/2019-59$. Authors are grateful to Dmitry Gorbunov for discussion.
\bibliographystyle{apsrev4-2}
\bibliography{refsR2.bib}
\end{document}